\documentclass[12pt]{article}
\usepackage{graphicx}

      \usepackage[cp1251]{inputenc} 
      \usepackage[russian]{babel}   

\begin{document}
 \noindent {\footnotesize\it Astronomy Letters, 2012 Vol. 38, No. 9, pp. 549--561}

 \noindent
 \begin{tabular}{llllllllllllllllllllllllllllllllllllllllllllll}
 & & & & & & & & & & & & & & & & & & & & & & & & & & & & & & & & & & & & & \\\hline\hline
 \end{tabular}

 \vskip 1.0cm
 \centerline {\large\bf Redetermination of Galactic Spiral Density Wave Parameters}
 \centerline {\large\bf  Based on Spectral Analysis of Maser Radial Velocities}
 \bigskip
 \centerline {A.T. Bajkova$^1$ and V.V. Bobylev$^{1,2}$}
 \medskip
{\small\it
 $^1$~Pulkovo Astronomical Observatory, Russian Academy of Sciences

 $^2$~Sobolev Astronomical Institute, St. Petersburg State University, Russia

 }

 \bigskip
{\bf Abstract}---To redetermine the Galactic spiral density wave
parameters, we have performed a spectral (Fourier) analysis of the
radial velocities for 44 masers with known trigonometric
parallaxes, proper motions, and line--of--sight velocities. The
masers are distributed in a wide range of Galactocentric distances
 $(3.5<R<13.2$~kpc) and are characterized by a wide scatter of position
angles $\theta$ in the Galactic $XY$ plane. This has required an
accurate allowance for the dependence of the perturbation phase
both on the logarithm of the Galactocentric distances and on the
position angles of the objects. To increase the significance of
the extraction of periodicities from data series with large gaps,
we have proposed and implemented a spectrum reconstruction method
based on a generalized maximum entropy method. As a result, we
have extracted a periodicity describing a spiral density wave with
the following parameters from the maser radial velocities: the
perturbation amplitude $f_R = 7.7^{+1.7}_{-1.5}$ km s$^{-1}$, the
perturbation wavelength $\lambda=2.2^{+0.4}_{-0.1}$~kpc, the pitch
angle of the spiral density wave
 $i=-5^{+0.2^\circ}_{-0.9^\circ}$,
 and the phase of the Sun in the
spiral density wave $\chi_\odot= -147^{+3^\circ}_{-17^\circ}$.

\section{Introduction}

A spectral analysis of the residual space velocities for various
young Galactic objects (HI clouds, OB stars, open clusters younger
than 50 Myr, masers) tracing the spiral arms was used, for
example, by Clemens (1985), Bobylev et al. (2008), and Bobylev and
Bajkova (2010). As a result, such spiral density wave parameters
(in accordance with the model of Lin and Shu (1964)) as the pitch
angle, the perturbation amplitude and wavelength, and the phase of
the Sun in the spiral density wave were determined. The spectral
analysis performed previously was the simplest periodogram
analysis (based on the Fourier transform) of the residual space
velocities for the objects as functions of their Galactocentric
distances. In this case, the position angles of the objects in the
Galactic $XY$ plane were disregarded.

Obviously, the previously applied approach may be considered only
as the first approximation that is accurate enough only when the
objects being analyzed occupy a comparatively small range of
Galactocentric distances. For example, in the case of OB stars and
open clusters localized within $\approx2-3$~kpc of the Sun, this
first approximation is quite adequate. In contrast, in the case of
using the currently available data on masers distributed in a wide
range of Galactocentric distances, $3<R<14$~kpc, a more accurate
processing that takes into account both the logarithmic dependence
of the perturbation phase on the Galactocentric distances and the
dependence of the perturbation phase on the position angles of the
objects is required. As will be shown below, our periodogram
analysis allows the significance of the extraction of
periodicities from measurements to be increased considerably.

Previously, we (Bobylev and Bajkova 2010) performed a spectral
analysis of the space velocities for 28 masers. At present, highly
accurate VLBI measurements are available already for 44 masers.
This is of great interest in redetermining the spiral density wave
parameters from objects of this class.

Thus, the goal of this paper is the development of a new, more
accurate approach to a periodogram analysis of the residual
velocities for Galactic objects and its application to redetermine
the Galactic spiral density wave parameters from masers
distributed in a wide range of Galactocentric distances. In
addition, to increase the significance of the extraction of
periodicities from data series with large gaps, we proposed and
implemented a method of analysis based on the reconstruction of
the spectra for nonuniform data series using a generalized maximum
entropy method (Bajkova 1992).

The paper is structured as follows. In the first section, we
consider the details of the developed periodicity extraction
method and present the results of its testing on model data. The
second section is devoted directly to an analysis of the radial
velocities for the largest number of masers known to date for
which highly accurate measurements of the trigonometric
parallaxes, proper motions, and line--of--sight velocities are
available.

 \section[]{THE METHOD}
 \label{MethodOfCalc}
 \subsection[]{Basic Relations}
 \label{basic}

The velocity perturbations of Galactic objects produced by a
spiral density wave (Lin and Shu 1964) are described by the
relations
\begin{equation}
V_R = - f_R \cos\chi,
\end{equation}
\begin{equation}
\Delta V_{\theta} = f_{\theta} \sin\chi,
\end{equation}
where
\begin{equation}
\chi = m[\cot(i)\ln(R/R_{\circ})-\theta]+\chi_{\odot}
\end{equation}
is the phase of the spiral density wave; $m$ is the number of
spiral arms; $i$ is the pitch angle; $\chi_{\odot}$ is the phase
of the Sun in the spiral density wave (Rohlfs 1977); $R_{\circ}$
is the Galactocentric distance of the Sun; $\theta$ is the
object’s position angle: $\tan\theta = y/(R_\circ-x)$, where $x,$
$y$ are the Galactic heliocentric rectangular coordinates of the
object; $f_R$ and $\Delta f_\theta$ are the amplitudes of the
radial and tangential perturbation components, respectively; $R$
is the distance of the object from the Galactic rotation axis,
which is calculated using the heliocentric distance $r=1/\pi$:
 $$
 R^2=r^2\cos^2 b-2R_\circ r\cos b\cos l+R^2_\circ,
 $$
where $l$ and $b$ are the Galactic longitude and latitude of the
object, respectively.

Equation (3) for the phase can be expressed in terms of the
perturbation wavelength $\lambda$, which is equal to the distance
between the neighboring spiral arms along the Galactic radius
vector. The following relation is valid:
\begin{equation}
\frac{2\pi R_{\circ}}{\lambda} = m\cot(i).
\end{equation}
Equation (3) will then take the form
\begin{equation}
\chi = \frac {2\pi R_{\circ}}{\lambda}
\ln(R/R_{\circ})-m\theta+\chi_{\odot}.
\end{equation}
Here, we will consider only the radial velocities and,
accordingly, the perturbations described by Eq.~(1), because
determining the residual tangential velocities with the needed
accuracy to study the perturbations~(2) is a rather complex task
(especially in the case of a small number of objects) that
requires constructing a smooth rotation curve with the highest
accuracy. In contrast, the radial velocities do not depend on the
rotation curve. The question of determining the residual
velocities is considered below. To goal of our spectral analysis
of the series of measured velocities $V_{R_n},$ $n=1,2,\dots,N,$
where $N$ is the number of objects, is to extract the periodicity
in accordance with model~(1) describing a spiral density wave with
parameters $f_R,$ $\lambda,$ and $\chi_\odot$. If the wavelength л
is known, then the pitch angle $i$ is easy to determine from
Eq.~(4) by specifying the number of arms m. Here, we adopt a
two-armed model, i.e., $m=2$.

\subsection{Spectral Analysis of the Perturbations}
\subsubsection{The linear approximation}

The linear approximation for the logarithm of the argument for
$|R-R_{\circ}|<<R_{\circ}$ and a small $\theta$ can be represented
as
\begin{equation}
\frac {2\pi R_{\circ}}{\lambda} \ln(R/R_{\circ})\approx \frac
{2\pi (R-R_{\circ})}{\lambda}.
\end{equation}
In this case, for our harmonic analysis of the velocities, we can
apply the standard Fourier transform
\begin{equation}
 \bar{V}_{\lambda_k} = \frac{1} {N}\sum_{n=1}^{N} V_{R_n}
 \exp\Bigl(-j\frac{2\pi}{\lambda_k}(R_n-R_{\circ})\Bigr),
\end{equation}
where $\bar{V}_{\lambda_k}$ is the $k$th harmonic of the Fourier
transform,  $V_{R_n}$ are the velocity measurements for objects
with Galactocentric distances $R_n,n=1,2,...,N$, and $\lambda_k$
is the wavelength of the $k-$th harmonic, which is equal to $D/k$,
where $D$ is the period of the series being analyzed.

Since we are interested only in the perturbation power spectrum
$|\bar{V}_{\lambda_k}|^2$, Eq.~(7) can be simplified as follows:
\begin{equation}
 \bar{V}_{\lambda_k} = \frac{1} {N}\sum_{n=1}^{N} V_{R_n}
 \exp\Bigl(-j\frac{2\pi}{\lambda_k}R_n\Bigr).
\end{equation}
We used the latter expression previously (see Bobylev et al. 2008;
Bobylev and Bajkova 2010) for our spectral analysis of the
residual velocities for Galactic objects.

\subsubsection{Analysis of the perturbations as functions
of the logarithm of the distances.}

Let us analyze the perturbations as a periodic function of the
logarithm of the Galactocentric distances, for the time being,
without allowance for the position angles of the objects:
\begin{equation}
 \bar{V}_{\lambda_k}=\frac{1} {N}\sum_{n=1}^{N} V_{R_n}
 \exp\Bigl(-j\frac {2\pi R_{\circ}}{\lambda_k}\ln(R_n/R_{\circ})\Bigr).
\end{equation}
Obviously, if we make the change of variables
\begin{equation}
 R^{'}_{n}=\ln(R_n/R_{\circ})R_{\circ},
\end{equation}
then Eq. (9) is reduced to the standard Fourier transform
\begin{equation}
 \bar{V}_{\lambda_k} = \frac{1} {N}\sum_{n=1}^{N} V_{R^{'}_n}
 \exp\Bigl(-j\frac {2\pi R^{'}_n}{\lambda_k}\Bigr).
\end{equation}

\begin{figure}[t]
{\begin{center}
 \includegraphics[width=0.99\textwidth]{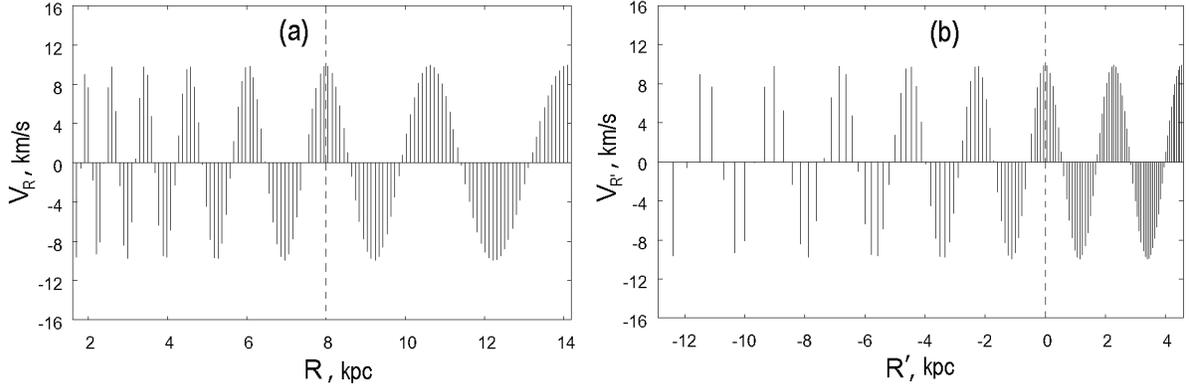}
 \caption{
Illustration of the change of variables when analyzing the
perturbations as a function of the logarithm of the distances.}
 \label{f1}
\end{center}}
\end{figure}

Figure 1 gives an illustration of this change, when the periodic
function of the logarithm of $R$ (Fig. 1a) turns into a periodic
function of the new variable $R'$ (Fig. 1b), which can already be
analyzed using the standard Fourier transform.

Below, in the ``Simulation Results'' Section, we will show how
important an accurate allowance for the logarithmic pattern of the
spiral density wave is, especially when the objects are
distributed in a wide range of Galactocentric distances.

\subsubsection{Allowance for the position angles of the objects}

Allowance for the position angles of the objects is a much more
complex algorithmic problem. Let us represent Eq.~(3) for the
phase as
\begin{equation}
 \chi = \chi_1-m\theta,
\end{equation}
where
\begin{equation}
 \chi_1 = \frac {2\pi R_{\circ}}{\lambda}\ln(R/R_{\circ})+\chi_{\odot}.
\end{equation}
Substituting (12) into Eq. (1) for the perturbations at the $n$th
point and performing standard trigonometric transformations, we
will obtain
 \begin{equation}
 \begin{array}{rll}
 &V_{R_n}=f_R \cos(\chi_{1_n} - m\theta_n)\\
 &      =f_R\cos\chi_{1_n}\cos m\theta_n   + f_R\sin\chi_{1_n}\sin m\theta_n \\
 &      =f_R \cos\chi_{1_n}(\cos m\theta_n + \tan\chi_{1_n}\sin  m\theta_n).
 \label{V-R-n}
 \end{array}
 \end{equation}
Let us designate
\begin{equation}
 V^{'}_{R}=f_R\cos\chi_1,
\end{equation}
Owing to the substitution (15), it then follows from (14) that
\begin{equation}
 V_{R_n}=V^{'}_{R_n}(\cos m\theta_n+\tan\chi_{1_n} \sin m\theta_n).
\end{equation}
Using Eq. (16), let us form a new data series
\begin{equation}
 V^{'}_{R_n}=V_{R_n}/(\cos m\theta_n+\tan\chi_{1_n} \sin m\theta_n),
\end{equation}
to which a Fourier analysis can be applied in accordance with
(11).

Thus, taking into account both the logarithmic pattern of the
spiral density wave and the position angles of the objects, we
obtain the following expression for our spectral analysis of the
perturbations:
\begin{equation}
 \bar{V}_{\lambda_k} = \frac{1} {N}\sum_{n=1}^{N} V^{'}_{R^{'}_n}
 \exp\Bigl(-j\frac {2\pi R^{'}_n}{\lambda_k}\Bigr).
\end{equation}

\subsubsection{The practical algorithm}

The algorithm for realizing (18) consists of the following steps:

{\bf1}.~The initial series of velocities $V_{R_n}$ is transformed
into the series $V_{R_n^{'}}$ in accordance with~(10).

{\bf2}.~The power spectrum of the derived sequence $V_{R_n^{'}}$
is calculated based on the Fourier transform~(11) to obtain a
estimate of $\lambda_{max}$ that corresponds to the peak of the
derived power spectrum.

{\bf3}.~A comb of several $\lambda_i (i=1,\dots,K)$ with a central
$\lambda_{max}$ is then specified.

The following iterations are made for each $\lambda_i$ from the
specified comb:

1)~The value of $\lambda_i$ and the initial approximation
$\chi_\odot$ (for example, equal to zero) are substituted into Eq.
(13) to calculate $\chi_1$ for each data reading ($n=1,\dots,N$).

2)~Using Eq. (17), the series of velocities $V_{R_n^{'}}$ is
transformed into the series $V^{'}_{R^{'}_n}$. This transformation
needs to be regularized to avoid the division by numbers close to
zero. This is done by assigning a threshold number, say, е, and
permission for the division is given only when the denominator in
Eq. (17) exceeds this number. The best $\varepsilon$ at which the
significance of the extracted peak in the spectrum reaches its
maximum as a result of the iterations at the minimum residual
between the solution and the data can be found by an exhaustive
search for е from some interval. The typical values of
$\varepsilon$ found on model problems lie within the range [0.01,
0.3].

3)~The power spectrum of the derived sequence $V^{'}_{R^{'}_n}$ is
calculated based on the Fourier transform~(18) to obtain a new
estimate of $\chi_\odot$ corresponding to a fixed $\lambda_i$ of
the derived power spectrum.

4)~The return to the first step is made until the process will
converge or diverge.

5)~If the process converged, then we fix the specified $\lambda_i$
and the derived $\chi_\odot$; if it diverged, then we take the
next value $\lambda_{i+1}$ from the specified comb and make
iterations (1)--(4) until the value of $\lambda$ at which the
process converges will be found.

{\bf4}.~The power spectrum is calculated for the values of
$\lambda$ and $\chi_\odot$ found in accordance with~(18) with the
goal of a further analysis.

The above algorithm is basically interactive and can be
efficiently applied to process the individual data realizations.
In contrast, in the case of mass data processing, for example,
during Monte Carlo simulations (see below), automation of the
search for periodicities is required. For this purpose, we propose
a slight modification of the algorithm described above that
consists in seeking for the best solution by maximizing some
criterion for the quality of signal extraction from noise. As such
a criterion, we propose a parameter $Q$ proportional to the peak
value of the power spectrum $S_{peak}$ and its significance
$p_{peak}$ (see the next subsection) and inversely proportional to
the value of the maximum side lobe in the power spectrum
$S_{sidelobe}$ and the residual between the extracted periodic
signal and the input data $\delta$, i.e., we suggest finding
$$
  \max~~Q=\frac{S_{peak}\times p_{peak}}{S_{sidelobe}\times \delta}
$$
by varying $\lambda_i (i=1,\dots,K)$ and $\varepsilon_j
(j=1,\dots,L)$.

As a result, as the solution we take the values of the parameters
$\lambda$, $f_R$ and $\chi_\odot$ at which $Q$ reaches its
maximum.  Note that the solution for $f_{R_i}$ for each
$\lambda_i$ is sought from the condition for the residual between
the extracted harmonic and the input data being at its minimum.

The high robustness of the constructed algorithm was established
through numerous simulations of the extraction of a harmonic
signal from noise at various numbers and various sampling
intervals of data, various signal-to-noise rations, and various
degrees of nonuniformity of the data series.

\subsubsection{The criterion for extracting a harmonic signal
from noise}

Since we actually solve the problem of extracting a harmonic
signal from noise, statistical criteria for separating the signal
and noise components should be applied. Here, we use a well-known
criterion based on Schuster’s theorem (Vityazev 2001). It consists
in specifying a positive number $q\ll1$ that defines the
probability (significance level) of a very rare event—the
appearance of a strong peak in the power spectrum (periodogram) of
white noise. In the case where the frequency of the periodic
component is not known in advance, it is quite natural to assume
that the largest value of the periodogram
$|\bar{V}_{\lambda_k}|^2_{max}$ corresponds to the sought-for
periodic component. If the following inequality holds:
$$
|\bar{V}_{\lambda}|^2_{max}\ge \frac{\sigma_0^2}{N}X,
$$
where
$$
 \renewcommand{\arraystretch}{1.4}
 \begin{array}{rll}
 \displaystyle
 \sigma_0^2&=& \displaystyle\frac{1}{N-1}\sum_{n=1}^{N}(V_{R_n}-\bar{V})^2,\\
 \bar{V}   &=& \displaystyle\frac{1}{N}\sum_{n=1}^{N}V_{R_n},\\
         X &=& \displaystyle-\ln(1-(1-q)^{2/(N-2)}),\\
 \label{sigma-0}
 \end{array}
$$
\renewcommand{\arraystretch}{1.2}
then the assertion that $|\bar{V}_{\lambda}|^2_{max}$ belongs to
the signal and not to the noise is adopted with the probability
 $p=1-q.$

\subsubsection{Monte Carlo simulations}

We use statistical Monte Carlo simulations to estimate the errors
of the parameters being determined. In accordance with this
method, we generate $M$ independent realizations of the velocities
and coordinates for the objects by taking into account their
random measurement errors that are known to us.

We assume that the data measurement errors are distributed
according to a normal law with a mean equal to the nominal value
and a dispersion equal to $\sigma_l={error}_l,$ $l=1,\dots,N_d$,
where $N_d$ is the number of data and ${error}_l$ designates the
measurement error of a single measurement with number $l$ (one
sigma). Each element of a random realization is formed
independently by adding the nominal value of the measured data
with number $l$ and a random number generated according to a
normal law with a zero mean and dispersion $\sigma_l.$ Note that
the latter is limited from above by $3\sigma_l$.

Subsequently, each random realization of data with number $j$
($j=1,\dots,M$) formed in this way is processed according to the
algorithm described above to determine the sought-for parameters
$f_R^j, \lambda^j, \chi_\odot^j$. The means of the parameters and
their dispersions are then determined from the derived sequences
of estimates: $m_{f_R}\pm \sigma_{f_R}, m_{\lambda}\pm
\sigma_{\lambda}, m_{\chi_\odot}\pm \sigma_{\chi_\odot}$. The
statistical characteristics of the spiral density wave pitch angle
$i$ can be easily determined using Eq.~(4): $m_{i}\pm \sigma_{i}$.

 \subsection{Spectrum Reconstruction by the Maximum Entropy Method}

So far we have considered the simplest method of periodogram
analysis for series. In the case where the data series are
irregular, i.e., there are large gaps, the signal spectrum is
distorted by large side lobes and it becomes difficult to
distinguish the spectral component of the signal from spurious
peaks. In this case, it may turn out to be useful to apply the
methods of spectrum reconstruction from the available data. This
problem is fundamentally resolvable if the sought for signal has a
finite spectrum. Since our problem belongs to the class of
problems on the extraction of polyharmonic functions from noise,
we assume that this condition is met.

There are two main nonlinear methods for reconstructing both
one-dimensional signals and images---these are the CLEAN method
and the maximum entropy method (MEM). Here, we consider the MEM as
a more fundamental method that has a rigorous logical
justification (Jaynes 1968). Since the spectrum, i.e., a
complex-valued function, is the function to be reconstructed, we
apply the generalized MEM (GMEM) described in detail by Bajkova
(1992) and Frieden and Bajkova (1994).

We will use the following notation:

1.The input data of the series: $V_{n} = V^{'}_{R^{'}_n},
n=1,\dots,N$, with coordinates $l_n$ on a discrete mesh
$1,\dots,K=2^\alpha$, $\alpha$ is an integer, $>0$;

2.The discrete Fourier spectrum of the input data: $X_k+jY_k$,
$k=1,\dots,K$.

The spectrum and the data are related by the inverse Fourier
transform
$$
 \frac{1} {K}\sum_{k=1}^{K} (X_k+jY_k)
 \exp\left(j\frac {2\pi (k-1) (l_n-1)}{K}\right) = V_{n}.
$$
Given the Hermitian symmetry of the spectrum for a real-valued
signal and the data measurement errors, the constraints on the
unknowns can be rewritten as
$$
 \sum_{k=K_1}^{k=K_2} X_k a_{k,n}-Y_k b_{k,n}+\eta_n =V_n,
$$
where
 $\displaystyle a_{k,n}=\frac{2}{K}\cdot\cos\left(\frac {2\pi (k-1)(l_n-1)}{K}\right)$,
 $\displaystyle b_{k,n}=\frac{2}{K}\cdot\sin\left(\frac {2\pi (k-1) (l_n-1)}{K}\right)$,
$\eta_n$ is the measurement error of the $n$th value of the series
that obeys a random law with a normal distribution with a zero
mean and dispersion $\sigma_n$.

In our case, the reconstruction problem consists in finding the
maximum of the following generalized entropy functional:
$$
 E=-\sum_{k=K_1}^{k=K_2}
 X_k^+\ln(aX_k^+)+X_k^-\ln(aX_k^-)+Y_k^+\ln(aY_k^+)+Y_k^-\ln(aY_k^-)-\sum_{n=1}^{n=N}\frac{\eta_n^2}{\sigma_n^2},
$$
where the sought--for variables $X_k$ and $Y_k$ are represented as
the difference of the positive and negative parts:
$X_k=X_k^+-X_k^-$ и $Y_k=Y_k^+-Y_k^-,$ respectively; in this case,
$X_k^+, X_k^-, Y_k^+, Y_k^-\ge0$, $a>0$ is the real--valued
parameter responsible for the separation of the positive and
negative parts of the sought-for variables with the required
accuracy (in our case, we adopted $a=1000)$, $K_1$ and $K_2$ are
the a priori known lower and upper localization boundaries of the
sought-for finite spectrum.

\subsection{Simulation Results}

In this section, we present the results of testing our developed
algorithms of spectral analysis based on both the Fourier
transform and the series reconstruction using the GMEM (Fig.~2).

The data series were formed in accordance with the relation
describing the spiral density wave (see also (1) and (5)):
$$
 V_{R_n}=f_R\cos\left(\frac{2\pi
 R_\circ}{\lambda}\ln(R_n/R_\circ)-m\theta_n+\chi_\odot\right)+\eta_n,~~n=1,\dots,N,
$$
where $\eta_n$ is additive white noise with a normal distribution
law with dispersion $\sigma$ and a zero mean.

We adopted the following model parameters: the perturbation
amplitude of the spiral density wave $f_R=10$ km s$^{-1}$; the
perturbation wavelength $\lambda=2.1$ kpc; the Galactocentric
distance of the Sun $R_\circ=8$ kpc; the number of spiral arms
$m=2$; the Sun’s phase in the spiral density wave
$\chi_\odot=90^\circ$; the number of objects $N=44$ (taken to be
equal to the number of Galactic masers whose velocities are
analyzed below); the position angles $\theta_n$ of the objects
were specified randomly in accordance with a uniform distribution
law within the range $[-\pi/4,\pi/4]$. The power spectrum of such
a signal in the absence of noise consists of a single peak with an
amplitude $f_R^2/4=25$ km$^2$ s$^{-2}$. The signal-to-noise ratio
of the model data was 2.63.

\begin{figure}[p]
{\begin{center}
 \includegraphics*[width=0.98\textwidth]{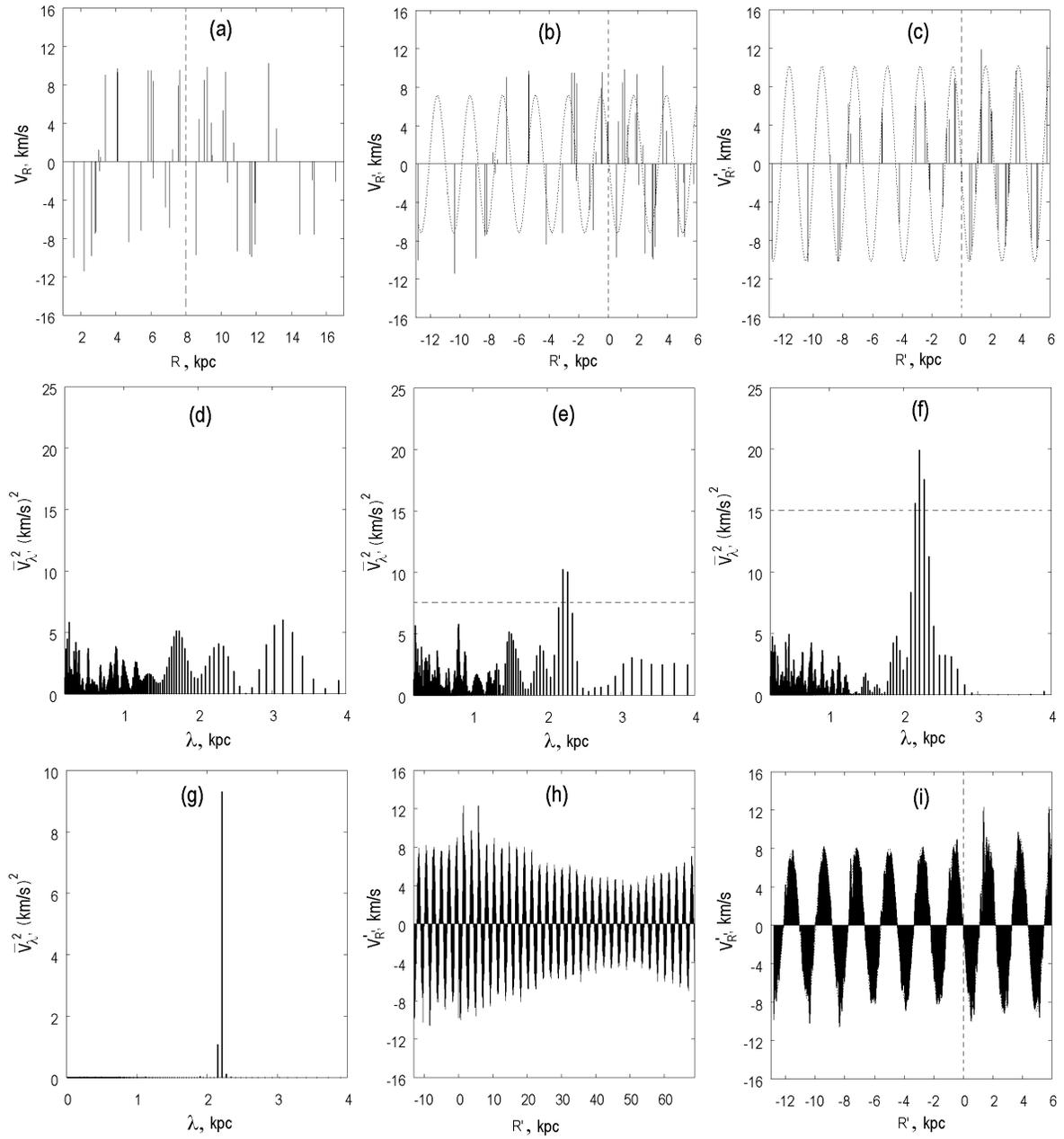}
 \caption{
Simulation results.}
 \label{f2}
\end{center}}
\end{figure}

The initial model sequence of radial velocities $V_{R_n}$ is shown
in Fig. 2a. It differs from the strictly harmonic one due to the
scatter of position angles for the objects and the measurement
errors. The range of Galactocentric distances is from 1.5 to 17
kpc. The sequence of velocities $V_{R^{'}_n}$ derived in
accordance with Eq. (10), which transforms the logarithmic
distances into the linear ones, is shown in Fig. 2b. The vertical
dashed line in the figures passes through the point corresponding
to the Galactocentric distance of the Sun.

To be able to apply fast computational algorithms (fast Fourier
transform), we represented our data as a discrete sequence on a
uniform mesh ${\rm\bf N}=2^{14}$ pixels in size and took the size
of a single pixel to be 0.005~kpc to provide data pixelization
with the required accuracy (in order that no more than one
measurement fall into one pixel) and to obtain the
highest--resolution spectrum. As a result, the length of the
analyzed period for the discrete sequence was $D={\rm\bf N}\times
0.005=81.92$~kpc. Obviously, the values of the N--point sequence
are taken to be zero in the pixels into which no data fall.

The power spectrum of the initial sequence $V_{R_n}$ calculated
(in accordance with the linear approximation of the logarithm of
the argument~(6)) using the Fourier transform~(8) is shown in
Fig.~2d. It is difficult to extract the significant peak
corresponding to the model signal from this spectrum.
Consequently, the linear approximation~(6) is not satisfactory in
our case.

The power spectrum of the sequence $V_{R^{'}_n}$ calculated from
(11) is shown in Fig. 2e. We see from this figure that taking into
account the logarithmic pattern of the spiral density wave (we
have not yet taken the maser position angles into account) allowed
a significant peak with an amplitude of~10 km$^2$ s$^{-2}$ to be
extracted at the required wavelength $\lambda=2.1$~kpc, which,
however, is a factor of 2.5 lower than the theoretical one. The
probability that the peak belongs to the signal and not to the
noise at the level indicated by the horizontal dashed line is
$p=0.95$. (The dashed line here and in similar figures is drawn at
3/4 of the maximum value of the periodogram.) The sought-for
harmonic of the perturbations corresponding to the peak and having
the smallest residual with the data is indicated by the dotted
line in Fig. 2b. Its amplitude is $f_R=7$ km s$^{-1}$. The Sun’s
phase in the spiral density wave is $\chi_\odot=81.2^\circ$.

Allowance for both the logarithmic pattern of the spiral density
wave and the position angles of the objects according to the
scheme described in the ``Practical algorithm'' Section led to the
result shown in Figs. 2c and 2f. It can be seen from Fig.~2c how
noticeably the envelope of the velocities changed once the
position angles had been taken into account: the values now fit
almost exactly into the harmonic signal indicated by the dashed
line with an amplitude $f_R=10$ km s$^{-1}$ and a wavelength
$\lambda=2.1$~kpc corresponding to the peak in the power spectrum
of the reconstructed signal (Fig.~2f). The amplitude of the peak
is 20 km$^2$ s$^{-2}$, which is twice that in the preceding case.
The significance of the peak at the level indicated by the
horizontal dashed line is $p=0.99997.$ The Sun's phase in the
spiral density wave is $\chi_\odot=92^\circ$.

To determine the range of possible solutions, i.e., the errors in
the parameters, we performed Monte Carlo simulations (see the
``Monte Carlo Simulations'' Section). We generated $M=1000$ random
realizations of data by varying their values within the limits of
measurement errors obeying a normal law. Having applied our method
of searching for periodicities to the simulated realizations of
data, we obtained the means and dispersions of the spiral density
wave parameters ($m_{par}\pm \sigma_{par}$): the perturbation
amplitude $f_R=7.4\pm 1.3$ km s$^{-1}$, the perturbation
wavelength $\lambda=2.2\pm 0.1$ kpc, the pitch angle of the spiral
density wave $i=-5.15^\circ\pm 0.15^\circ$, and the Sun’s phase in
the spiral density wave $\chi_{\odot}=87^\circ\pm 20^\circ$. Note
that the means of the parameters turned out to be slightly shifted
relative to the values corresponding to the nominal data.

Thus, using the proposed iterative scheme of periodogram analysis,
we managed to reconstruct the spiral density wave parameters from
the model radial velocities of 44 objects having a wide scatter in
both Galactocentric distances and position angles with a
sufficiently high accuracy. Clearly, the accuracy of determining
the spiral density wave parameters increases with increasing
number of objects and increasing accuracy of measuring their
coordinates and velocities.

As has already been said in the previous section, large side lobes
can be obtained in the case of large gaps in the data, which can
make it difficult to separate the signals of real and spurious
peaks in the periodogram. A spectral analysis with the application
of series reconstruction methods makes it possible to increase the
significance of extracting the useful signal. The reconstruction
results based on the GMEM as applied to the $V^{'}_{R^{'}_n}$,
data obtained by applying an iterative periodogram analysis are
shown in Figs. 2g, 2h, and 2i. Figures 2h and 2i show the
reconstructed sequence $V^{'}_{R^{'}_n}$ on the entire analyzed
period and in the region of data from $-12.82< R^{'}<5.8$ kpc. It
can be seen from Fig.~2g, where the reconstructed spectrum is
shown, that we managed to get rid of the side lobes seen in Figs.
2f almost completely and, thus, to increase the significance of
extracting the periodic signal $(p=1),$ determining its
wavelength, and the Sun's phase in the spiral density wave. Since
we failed to reconstruct the signal outside the range containing
the data (Fig.~2h) with a high accuracy, this led to a decrease in
the amplitude of the spectral peak in Fig.~2g. Nevertheless, we
managed to obtain accurate values of such parameters as the
wavelength $\lambda=2.1$ kpc and the Sun's phase
$\chi_\odot=90^\circ$.

\section{ANALYSIS OF MASERS}
\subsection{Data}

Previously (Bobylev and Bajkova 2010; Stepanishchev and Bobylev
2011), we analyzed a sample of 28 masers with measured
trigonometric parallaxes, proper motions, and line-of-sight
velocities drawn from published data. By now, the amount of such
data has increased considerably---about 20 new measurements in
various regions of active star formation have been published. The
initial data on 44 masers associated with the youngest Galactic
stellar objects (either protostellar objects of various masses, or
very massive supergiants, or T Tau stars) are given in the table.

The observational data, namely the trigonometric parallaxes and
proper motions of the objects, were obtained by several research
groups through longterm radio-interferometric observations within
the framework of various projects. One of them is the Japanese
VERA (VLBI Exploration of Radio Astrometry) project on the
observation of Galactic H2O masers at 22 GHz and SiO masers (there
are very few such sources among young objects) at 43 GHz. Note
that the higher the frequency, the higher the resolution, the more
accurate the observations. Methanol (CH3OH) masers are observed at
12 GHz on the VLBA (NRAO). The radio-interferometric observations
of radio stars in continuum at 8.4 GHz are being carried out with
the same goals.

\begin{figure}[t]
{\begin{center}
 \includegraphics[width=0.45\textwidth]{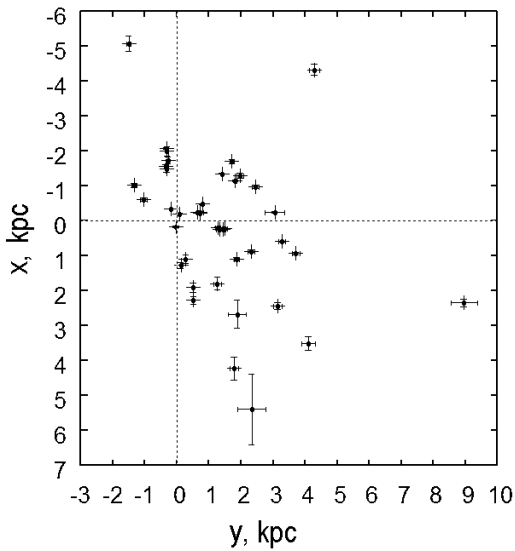}
 \caption{
Maser positions in projection onto the Galactic $xy$ plane (the
Sun is at the coordinate origin).}
 \label{f3}
\end{center}}
\end{figure}

\begin{figure}[p]
{\begin{center}
 \includegraphics[width=0.98\textwidth]{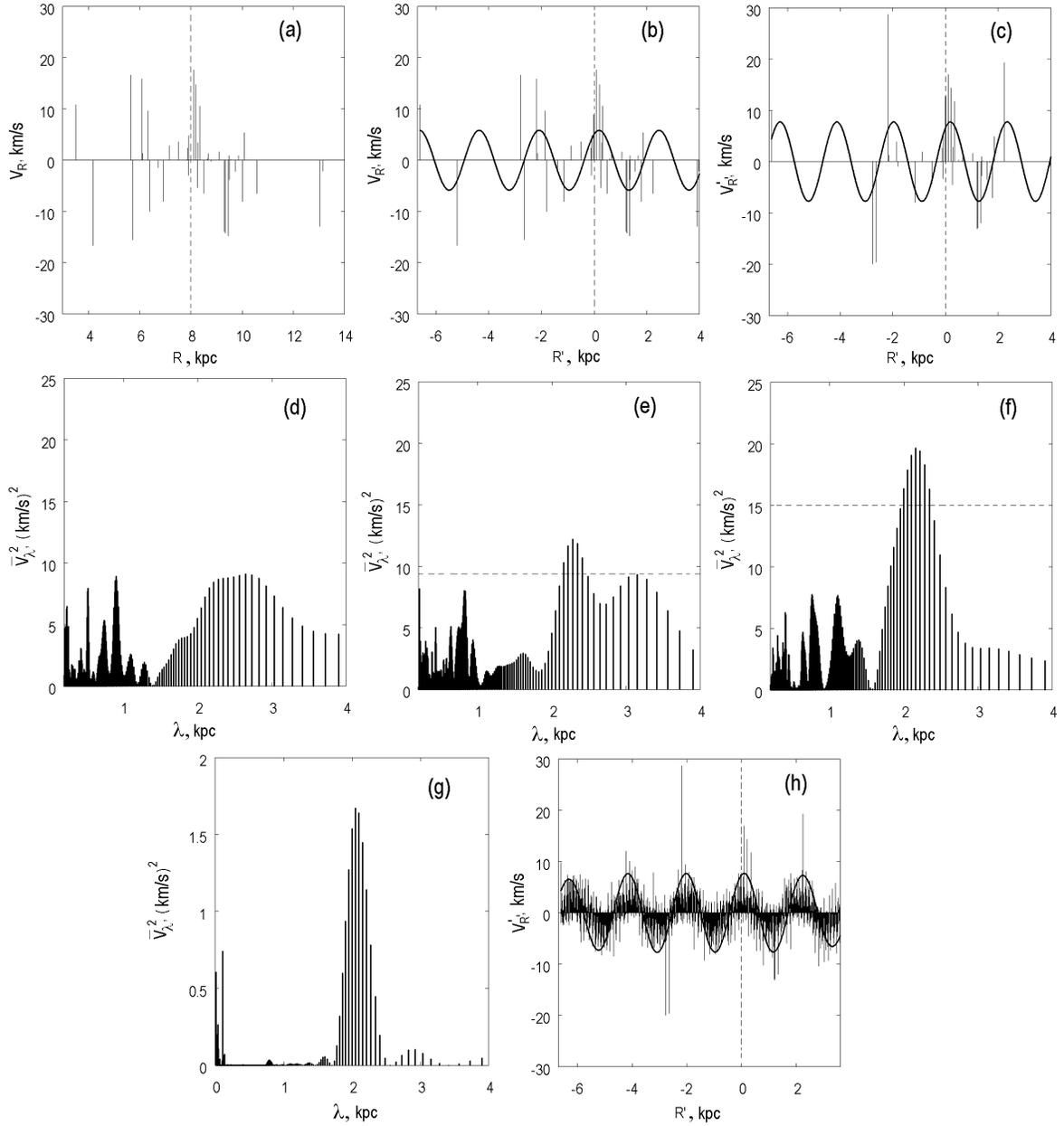}
  \caption{Results of processing the data on 44 masers.}
 \label{f4}
\end{center}}
\end{figure}

Several radio-interferometric determinations of the parallaxes and
proper motions have been made for a number of objects in the Orion
Nebula. These include SiO masers (Kim et al. 2008), H2O masers
(Hirota et al. 2007), radio observations of the radio star GMR A
(a low-mass T Tau star) at 15 GHz (Sandstrom et al. 2007), and
independent observations of several radio stars at 8.4 GHz (Menten
et al. 2007). Menten et al. (2007) deduced the mean parallax and
proper motion from four radio stars in the cluster of the Orion
Nebula: GMR A, GMR 12, GMR F, and GMR G. In the opinion of Kim et
al. (2008), the SiO masers are associated with the accretion disk
rotating around the source ``N'' in the central region of
Orion/KL. According to Goddi et al. (2011), this radio source is a
binary system with a total mass of $\approx20M_\odot$ Note that
the proper motions from different measurements differ
significantly. Therefore, we used two independent measurements
indicated in the table.

The star YV CMa, a red supergiant with a mass of
$\approx25M_\odot$ and an age of $\approx8$~Myr, is surrounded by
a thick expanding envelope (Zhang et al. 2012). The red supergiant
S Per is also surrounded by an expanding envelope (Asaki et al.
2010).

In the star-forming region G5.89--0.39, the water masers that were
used to determine the trigonometric parallax are associated with
the accretion disk located inside an ultracompact HII region,
which is at the expansion stage itself (Motogi et al. 2011). The
gas is ionized by a just born massive $\approx$O8 star. An
overview of the complex picture in this region can be found in Xu
et al. (2012).

In the star-forming region Cep A, we use both observations of
methanol masers at 12~GHz (Moscadelli et al. 2009) for the source
HW2 (a massive protostellar object) and independent continuum
observations of the radio source HW9 (this is either a T tau star
or a young Ae/Be star with a mass of less than $6M_\odot)$ at 8.4
GHz (Dzib et al. 2011).

In the star-forming region NGC 1333, we averaged the proper
motions of two maser features, f1 and f2, that are associated with
the young stellar object SVS 13 (Hirota et al. 2008a). New data
(Hirota et al. 2011) on the young stellar object with clear
evidence of a bipolar jet have been obtained in this region
associated with the molecular cloud Lynds 1448 C in Perseus.

The line-of-sight velocities $V_r~(LSR)$ in the table are given
relative to the Local Standard of Rest. Their values were
determined by different authors from radio observations in CO
lines. Occasionally, the cited authors do not provide the error in
the line-of-sight velocity; we then took it to be $5$~km s$^{-1}$.

As follows from the table, on average, the parallaxes were
determined with a relative error $\sigma_\pi/\pi\approx5\%,$ and
only in three regions does it exceed appreciably the mean level.
These are IRAS 16293-2422($\sigma_\pi/\pi=19\%$),
 G~23.43-0.20 ($\sigma_\pi/\pi=18\%$) and
 W~48 ($\sigma_\pi/\pi=14\%$).

The distribution of masers in projection onto the Galactic $xy$
plane is shown in Fig.~3, from which we see a fairly wide scatter
in coordinates $x,y$ and, hence, in position angles in the
Galactic $xy$ plane.

      \begin{table}[]
      \begin{center}
      \caption{Initial data on the masers}
      {\small
      \begin{tabular}{|l|r|r|r|r|r|r|r|r|r|r|r|}      \hline
      Source & $\alpha$ & $\delta$ & $\pi(\sigma_\pi)$ &
      $\mu^*_\alpha (\sigma_{\mu_\alpha})$ & $\mu_\delta(\sigma_{\mu_\delta})$ &
      $V_r(\sigma_{V_r})$ & Ref \\\hline

 S252A      & $ 92.2222$ & $ 21.6414$ & $ .476( .006)$ & $   .02( .01)$ & $ -2.02( .04)$ & $ 10.8( 3)$ & (1) \\\hline 
 G232.6+0.99& $113.0408$ & $-16.9702$ & $ .596( .035)$ & $ -2.17( .06)$ & $  2.09( .46)$ & $ 22.8( 3)$ & (1) \\\hline 
 Cep A      & $344.0754$ & $ 62.0304$ & $1.430( .080)$ & $   .50( 1.1)$ & $ -3.70( .20)$ & $-10.5( 5)$ & (1) \\\hline 
 NGC 7538   & $348.4390$ & $ 61.4696$ & $ .378( .017)$ & $ -2.45( .03)$ & $ -2.44( .06)$ & $-57.0( 3)$ & (1) \\\hline 
 V645       & $295.7969$ & $ 23.7343$ & $ .463( .020)$ & $ -1.65( .03)$ & $ -5.12( .07)$ & $ 27.4( 3)$ & (1) \\\hline 
 G35.20-0.74& $284.5544$ & $  1.6766$ & $ .456( .045)$ & $  -.18( .08)$ & $ -3.63( .16)$ & $ 27.9( 3)$ & (1) \\\hline 
 W48        & $285.4397$ & $  1.2257$ & $ .306( .045)$ & $  -.71( .07)$ & $ -3.61( .24)$ & $ 41.9( 3)$ & (1) \\\hline 
 G23.43-0.20& $278.6633$ & $ -8.5237$ & $ .170( .032)$ & $ -1.93( .10)$ & $ -4.11( .07)$ & $ 97.6( 3)$ & (1) \\\hline 
 G23.01-0.41& $278.6678$ & $ -9.0107$ & $ .218( .017)$ & $ -1.72( .04)$ & $ -4.12( .30)$ & $ 81.5( 3)$ & (1) \\\hline 
 Orion      & $ 83.8098$ & $ -5.3773$ & $2.425( .035)$ & $  3.30(1.0 )$ & $   .10(1.0 )$ & $ 10.0( 5)$ & (2) \\\hline 
 Orion/KL   & $ 83.8104$ & $ -5.3751$ & $2.390( .030)$ & $  9.56( .10)$ & $ -3.83( .15)$ & $  5.0( 5)$ & (3) \\\hline 
 W3 (OH)    & $ 36.7702$ & $ 61.8735$ & $ .512( .007)$ & $ -1.20( .02)$ & $  -.15( .01)$ & $-44.2( 3)$ & (4) \\\hline 
 IRAS 00420 & $ 11.2433$ & $ 55.7799$ & $ .460( .010)$ & $ -2.52( .05)$ & $  -.84( .04)$ & $-46.0( 5)$ & (5) \\\hline 
 IRAS 16293 & $248.0952$ & $-24.4768$ & $5.6  (1.1  )$ & $-20.6 ( .7 )$ & $-32.4 (2.0 )$ & $  4.4( 5)$ & (6) \\\hline 
 NGC 1333   & $ 52.2655$ & $ 31.2677$ & $4.250( .320)$ & $ 14.25(1.0 )$ & $ -8.95(1.4 )$ & $  7.5( 5)$ & (7) \\\hline 
 IRAS 22198 & $335.3614$ & $ 63.8605$ & $1.309( .047)$ & $ -3.00( .50)$ & $   .0 (1.0 )$ & $-17.0( 5)$ & (8) \\\hline 
 S269       & $ 93.6544$ & $ 13.8267$ & $ .189( .008)$ & $  -.42( .01)$ & $  -.12( .04)$ & $ 19.6( 3)$ & (9) \\\hline 
 WB89-437   & $ 40.8690$ & $ 62.9523$ & $ .164( .006)$ & $ -1.27( .05)$ & $   .82( .12)$ & $-72.0( 3)$ &(10) \\\hline 
 L1287      & $  9.1973$ & $ 63.4839$ & $1.077( .039)$ & $  -.86( .11)$ & $ -2.29( .56)$ & $-23.5( 3)$ &(11) \\\hline 
 NGC 281-W  & $ 13.1008$ & $ 56.5620$ & $ .421( .022)$ & $ -2.69( .16)$ & $ -1.77( .11)$ & $-29.5( 3)$ &(11) \\\hline 
 S255       & $ 93.2251$ & $ 17.9898$ & $ .628( .027)$ & $  -.14( .54)$ & $  -.84(1.76)$ & $  4.6( 3)$ &(11) \\\hline 
 L1206      & $337.2142$ & $ 64.2281$ & $1.289( .153)$ & $   .27( .23)$ & $ -1.40(1.95)$ & $-12.0( 3)$ &(11) \\\hline 
 S Per      & $ 35.7155$ & $ 58.5865$ & $ .413( .017)$ & $  -.49( .23)$ & $ -1.19( .20)$ & $-38.5( 1)$ &(12) \\\hline 
 IRAS 06061 & $ 92.2791$ & $ 21.8448$ & $ .496( .031)$ & $  -.10( .10)$ & $ -3.91( .07)$ & $ -1.6( 5)$ &(13) \\\hline 
 G14.33-0.64& $274.7278$ & $-16.7973$ & $ .893( .101)$ & $   .95(2.0 )$ & $ -2.50(2.0 )$ & $ 22.0(10)$ &(14) \\\hline 
 W51 Main/S & $290.9328$ & $ 14.5082$ & $ .185( .010)$ & $ -2.64( .16)$ & $ -5.11( .16)$ & $ 58.0( 4)$ &(15) \\\hline 
 IRAS 06058 & $ 92.2229$ & $ 21.6418$ & $ .569( .034)$ & $  1.06( .10)$ & $ -2.77( .20)$ & $  3.0( 3)$ &(16) \\\hline 
 IRAS 19213 & $290.9055$ & $ 17.4862$ & $ .251( .010)$ & $ -2.53( .10)$ & $ -6.07( .30)$ & $ 41.7( 3)$ &(16) \\\hline 
 AFGL 2789  & $324.9928$ & $ 50.2392$ & $ .326( .032)$ & $ -2.20( .13)$ & $ -3.77( .36)$ & $-44.0( 3)$ &(16) \\\hline 
 L1448C     & $ 51.4120$ & $ 30.7348$ & $ 4.31( .33 )$ & $ 21.90( .07)$ & $-23.10( .33)$ & $  4.5( 5)$ &(17) \\\hline 
 G 5.89-0.39& $270.1263$ & $-24.0679$ & $ .780( .050)$ & $   .17( .10)$ & $  -.95( .10)$ & $ 10.0( 3)$ &(18) \\\hline 
 Onsala 1   & $302.5383$ & $ 31.5267$ & $ .404( .017)$ & $ -3.10( .18)$ & $ -4.70( .24)$ & $ 12.0( 1)$ &(19) \\\hline 
 Onsala 2N  & $305.4334$ & $ 37.6104$ & $ .261( .009)$ & $ -2.79( .13)$ & $ -4.66( .17)$ & $   .0( 1)$ &(20) \\\hline 
      \end{tabular}}
      \end{center}
      \end{table}
      \begin{table*}[]
      \begin{center}
      Table 1 (Continue)
      {\small
      \begin{tabular}{|l|r|r|r|r|r|r|r|r|r|r|r|}      \hline
      Source & $\alpha$ & $\delta$ & $\pi(\sigma_\pi)$ &
      $\mu^*_\alpha (\sigma_{\mu_\alpha})$ & $\mu_\delta(\sigma_{\mu_\delta})$ &
      $V_r(\sigma_{V_r})$ & Ref \\\hline
 G192.1-3.8 & $ 89.5564$ & $ 16.5330$ & $ .660( .040)$ & $   .69( .15)$ & $ -1.57( .15)$ & $  5.7( 3)$ &(21) \\\hline 
 G12.9+0.45 & $272.9642$ & $-17.5250$ & $ .428( .022)$ & $   .16( .03)$ & $ -1.90(1.59)$ & $ 39.8( 5)$ &(22) \\\hline 
 M17        & $275.1034$ & $-16.1931$ & $ .505( .033)$ & $   .68( .05)$ & $ -1.42( .09)$ & $ 23.4( 5)$ &(22) \\\hline 
 G75.3+1.32 & $304.0667$ & $ 37.5961$ & $ .108( .005)$ & $ -2.37( .09)$ & $ -4.48( .17)$ & $-57.0( 2)$ &(23) \\\hline 
 W75N       & $309.6518$ & $ 42.6263$ & $ .772( .042)$ & $ -1.97( .10)$ & $ -4.16( .15)$ & $  9.0( 5)$ &(24) \\\hline 
 DR21       & $309.7529$ & $ 42.3806$ & $ .666( .035)$ & $ -2.84( .15)$ & $ -3.80( .22)$ & $ -3.0( 5)$ &(24) \\\hline 
 DR20       & $309.2540$ & $ 41.5821$ & $ .687( .038)$ & $ -3.29( .13)$ & $ -4.83( .26)$ & $ -3.0( 5)$ &(24) \\\hline 
 IRAS 20290 & $307.7111$ & $ 41.0410$ & $ .737( .062)$ & $ -2.84( .09)$ & $ -4.14( .54)$ & $ -1.4( 5)$ &(24) \\\hline 
 AFGL 2591  & $307.7111$ & $ 41.0410$ & $ .300( .010)$ & $ -1.21( .32)$ & $ -4.80( .12)$ & $ -5.7( 5)$ &(24) \\\hline 
 HW9 CepA   & $344.0777$ & $ 62.0300$ & $1.43 ( .07)$  & $  -.76( .11)$ & $ -1.85( .04)$ & $-10.5( 5)$ &(25) \\\hline 
 VY CMa     & $110.7430$ & $-25.7675$ & $ .830( .080)$ & $ -2.80( .20)$ & $  2.60( .20)$ & $ 18.0( 3)$ &(26) \\\hline 
      \end{tabular}}
      \end{center}
{\small
 Note. $\alpha$ and $\delta$ in degrees,
 $\pi$ is in mas,
 $\mu^*_\alpha=\mu_\alpha\cos\delta$ and $\mu_\delta$ is in mas yr$^{-1}$,
 $V_r=V_r(LSR)$ is in km s$^{-1}$; the numbers mark
the references to papers: (1) Reid et al. (2009); (2) Menten et
al. (2007); (3) Kim et al. (2008); (4) Xu et al. (2006); (5)
Moellenbrock et al. (2009); (6) Imai et al. (2007); (7) Hirota et
al. (2008a); (8) Hirota et al. (2008b); (9) Honma et al. (2007);
(10) Hachisuka et al. (2009); (11) Rygl et al. (2010); (12) Asaki
et al. (2010); (13) Niinuma et al. (2011); (14) Sato et al.
(2010a); (15) Sato et al. (2010b); (16) Oh et al. (2010); (17)
Hirota et al. (2011); (18) Motogi et al. (2011); (19) Nagayama et
al. (2011); (20) Ando et al. (2011); (21) Shiozaki et al. (2011);
(22) Xu et al. (2011); (23) Sanna et al. (2012); (24) Rygl et al.
(2011); (25) Dzib et al. (2011); (26) Zhang et al. (2012).
      }
      \end{table*}

\subsection{Results and Discussion}

First of all, we redetermined the parameters of the Galactic
rotation curve from the data on 44 masers. Our technique, which is
based on the expansion of the angular velocity of Galactic
rotation in a Taylor series of the Galactocentric distance $R,$
was described in detail previously (see Bobylev and Bajkova 2010;
Stepanishchev and Bobylev 2011).

At fixed $R_0=8$~kpc, we obtained the components of the peculiar
solar velocity
 $(U_\odot,V_\odot,W_\odot)=(7.6,17.8,8.3)\pm(1.5,1.4,1.2)$~km s$^{-1}$ and the following
Galactic rotation parameters:
 $\Omega_0=-28.8 \pm0.8$~km s$^{-1}$ kpc$^{-1}$,
 $\Omega'_0=+4.18\pm0.15$~km s$^{-1}$ kpc$^{-2}$,
 $\Omega''_0=-0.87\pm0.06$~km s$^{-1}$ kpc$^{-3}$.
The linear Galactic rotation velocity at $R=R_{\circ}$ is then
$V_0=|R_0\Omega_0|=230\pm14$~km s$^{-1}$.

There is good agreement of our results with the results of
analyzing masers by different authors. Based on a sample of 18
masers, McMillan and Binney (2010) showed that $\Omega_0$ lying
within the range $29.9-31.6$~km s$^{-1}$ kpc$^{-1}$ at various
$R_0$ was determined most reliably and obtained an estimate of
$V_0=247\pm19$~km s$^{-1}$ for $R_0=7.8\pm0.4$~kpc. Based on a
sample of 18 masers, Bovy et al. (2009) found $V_0=244\pm13$~km
s$^{-1}$ at $R_0=8.2$~kpc. Using 28 masers, Stepanishchev and
Bobylev (2011) found the following parameters:
 $(U_\odot,V_\odot,W_\odot)=(8.5,17.1,8.9)\pm(1.6,1.6,1.6)$~km s$^{-1}$, и
 $\Omega_0 =-30.4\pm0.7$~km s$^{-1}$ kpc$^{-1}$,
 $\Omega'_0= 4.23\pm0.13$~km s$^{-1}$ kpc$^{-2}$,
 $\Omega''_0=-1.01\pm0.06$~km s$^{-1}$ kpc$^{-3}$.

It is important that the rotation-curve parameters found are in
good agreement with the results of analyzing young Galactic disk
objects rotating most rapidly around the center: OB associations
with $\Omega_0 =-31\pm1$~km s$^{-1}$ kpc$^{-1}$ (Mel’nik et al.
2001; Mel’nik and Dambis 2009), blue supergiants with
$\Omega_0=-29.6\pm1.6$~km s$^{-1}$ kpc$^{-1}$ and
   $\Omega'_0= 4.76\pm0.32$~km s$^{-1}$ kpc$^{-2}$ (Zabolotskikh et al. 2002),
   or OB3 stars with
 $\Omega_0 = -31.5\pm0.9$~km s$^{-1}$ kpc$^{-1}$,
 $\Omega^{'}_0 = +4.49\pm0.12$~km s$^{-1}$ kpc$^{-2}$ and
 $\Omega^{''}_0 = -1.05\pm0.38$~km s$^{-1}$ kpc$^{-3}$ (Bobylev and Bajkova 2011).

The Galactocentric radial, $V_{R_n},$ and tangential,
$V_{\theta_n}$ ($n=1,\dots,44,)$ velocities of the masers were
determined from the relations
 \begin{equation}
  V_{\theta_n}= U_n\sin \theta_n+(V_0+V_n)\cos \theta_n,
 \end{equation}
\begin{equation}
  V_{R_n}=-U_n\cos \theta_n+(V_0+V_n)\sin \theta_n,
\end{equation}
where  $U_n, V_n$ are the heliocentric space velocities freed from
the peculiar solar velocity $U_\odot,V_\odot$ found.

The residual tangential velocities are obtained from the
tangential velocities (19) minus the smooth rotation curve that is
defined by the Galactic rotation parameters $\Omega_0,$
$\Omega'_0,$ and $\Omega''_0$ found. The radial velocities (20)
depend only on one Galactic parameter $\Omega_0$ and do not depend
on the rotation curve. As our experience showed (Bobylev and
Bajkova 2010), the data are so far insufficient to reliably
extract the density wave from the tangential residual velocities
of the masers. Therefore, here we determine the spiral density
wave parameters only from the radial velocities.

The results of our processing of the radial velocities are
presented in Fig. 4. The length of the discrete sequence and the
pixelization parameters were chosen to be the same as those in the
case of our simulations (see the ``Simulation Results'' Section).
Figure 4a shows the initial sequence of velocities $V_{R_n},
n=1,\dots,44$. We see that the Galactocentric distances occupy a
fairly wide range, from 3.5 to 13.2 kpc. The sequence of
velocities $V_{R^{'}_n}$ derived in accordance with Eq. (10),
which transforms the logarithmic distances into the linear ones,
is shown in Fig. 4b. The power spectrum of the initial sequence
$V_{R_n}$ that we calculated based on the Fourier transform (8) is
shown in Fig. 4d. We see from this figure that at least four peaks
have comparable significances and none of them can be extracted as
the main one. The peak near 2 kpc is too blurred to determine the
corresponding wavelength with an acceptable accuracy. Hence it
follows that it is inappropriate to use the linear approximation
(6) to analyze our data.

The power spectrum of the transformed sequence $V_{R^{'}_n}$,
calculated using (11) is shown in Fig. 4e. We see from this figure
that taking into account the logarithmic pattern of the spiral
density wave allowed a significant $(p = 0.83)$ peak equal to 12.2
km$^2$ s$^{-2}$. The harmonic corresponding to this peak and
having the smallest residual with the data is indicated in Fig. 4b
by the solid line. Its amplitude is $f_R=5.8$ km s$^{-1}$. The
Sun’s phase in the spiral density wave measured from the center of
the Carina–Sagittarius arm (Rohlfs 1977) ($R\approx 7$ kpc) is
$\chi_\odot=-149^\circ$.

Allowance for both the logarithmic pattern of the spiral density
wave and the position angles of the objects according to the
scheme described in the ``Practical algorithm'' Section led to the
result shown in Figs.~4c and 4f. It can be seen from Fig. 4c that
the sequence of velocities $V^{'}_{R^{'}_n}$ was modified
noticeably compared to $V_{R^{'}_n}$ in Fig.~4b, fitting more
closely into the harmonic signal (solid line) corresponding to the
largest peak in the power spectrum (Fig.~4f). The perturbation
amplitude is $f_R=7.7$ km s$^{-1}$ and the wavelength is
$\lambda=2.2$ kpc. The Sun’s phase in the spiral density wave is
$\chi_\odot=-147^\circ$. The amplitude of the peak in the power
spectrum is 20 km$^2$ s$^{-2}$, which is almost twice as high as
that in the previous case. The significance of the peak at the
level indicated by the horizontal dashed line in Fig.~4f is
$p=0.98,$ while the significance level of the peak in the spectrum
(Fig.~4e) is only $p=0.83.$

The results of our reconstruction using the GMEM as applied to the
$V^{'}_{R^{'}_n}$ data obtained during an iterative periodogram
analysis are shown in Figs.~4g and 4h. Figure 4g presents the
reconstructed power spectrum, while Fig. 4h presents the
reconstructed data series together with the extracted periodic
signal (solid line) corresponding to the peak in the power
spectrum and constituting the minimum residual with the initial
$V^{'}_{R^{'}_n}$. As we see from Fig.~4g, using the GMEM-based
spectrum reconstruction method allowed us to get rid of the side
lobes near the main peak almost completely and, thus, to increase
considerably the significance of the extracted periodicity $(p=1)$
with $\lambda=2.2$ kpc. The discrepancy between the extracted
periodic signal and the reconstructed sequence $V^{'}_{R^{'}_n}$
(Fig. 4h) is high-frequency noise in the wavelength range
$0<\lambda<0.11$~kpc (see Fig.~4g), which is of no interest for
our problem.

Based on our Monte Carlo simulations of 1000 random data
realizations by varying their values within the limits of
measurement errors obeying a normal distribution law, we obtained
the means and dispersions of the spiral density wave parameters
($m_{par}\pm\sigma_{par}$): the perturbation amplitude
$f_R=7.8\pm1.6$ km s$^{-1}$, the perturbation wavelength
$\lambda=2.35\pm0.25$ kpc, the pitch angle of the spiral density
wave $i=-5.36^\circ\pm0.57^\circ$ and the Sun’s phase in the
spiral density wave $\chi_{\odot}=-154^\circ\pm10^\circ$. Here, as
in the case of solving the model problem (the ``Simulation
Results'' Section), we obtained shifted means of the parameters
relative to the solutions obtained at the nominal values of the
input data. As a result, we ultimately have the following spiral
density wave parameters: the perturbation amplitude
$f_R=7.7^{+1.7}_{-1.5}$ km s$^{-1}$, the perturbation wavelength
$\lambda=2.2^{+0.4}_{-0.1}$ kpc, the pitch angle of the spiral
density wave $i=-5^{+0.2^\circ}_{-0.9^\circ}$ and the Sun’s phase
in the spiral density wave
$\chi_\odot=-147^{+3^\circ}_{-17^\circ}$.

Our values of such parameters as the perturbation amplitude $f_R,$
the wavelength $\lambda$, and the pitch angle of the spiral
density wave $i$ are in good agreement with the results of other
authors obtained by analyzing young Galactic disk objects. For
example, Mel’nik et al. (2001) found
 $f_R=-7\pm1$~km s$^{-1}$, $f_\theta=2\pm1$~km s$^{-1}$, and
 $\lambda=2.0\pm0.2$~kpc for $m=2$ by analyzing OB
associations. Zabolotskikh et al. (2002) found
 $f_R=-7\pm2$~km s$^{-1}$ and
 $f_\theta=-1\pm2$~km s$^{-1}$,
 $i=-6.0\pm0.9^\circ$ for $m=2$
 with a phase $\chi_\odot\approx-85^\circ$
from the data on young Cepheids ($P>9^d$) and open star clusters
($\log T<7.6$);
 $f_R=-6.6\pm2.5$~km s$^{-1}$ and
 $f_\theta=0.4\pm2$~km s$^{-1}$,
 $i=-6.6\pm0.9^\circ$ for $m=2$ with a phase $\chi_\odot\approx-97^\circ$
from the data on OB stars. Bobylev and Bajkova (2011) found
 $f_R=-12.5\pm1.1$~km s$^{-1}$,
 $f_\theta= 2.0\pm1.6$~km s$^{-1}$
(the signs at the amplitudes in the cited papers were reconciled
with Eqs. (1) and (2)),
 $i=-5.3\pm0.3^\circ$ for $m=2$ with a phase
 $\chi_\odot=-91\pm4^\circ$ from the data
on OB3 stars with an independent distance scale determined from
interstellar Ca~II absorption lines.

Nevertheless, we obtained a fairly paradoxical value,
$\chi_\odot=-147^\circ$. For young objects, which the masers are,
one might expect a phase $\chi_\odot\approx-90^\circ$. Note that
analysis of Cepheids yielded $\chi_\odot=-165\pm1^\circ$ (with the
phase measured from the Carina–Sagittarius arm) (Byl and Ovenden
1978), $\chi_\odot=-150^\circ$ was found from red supergiants and
Cepheids (Mishurov et al. 1979), or
$\chi_\odot\approx-290\pm16^\circ$ (Mishurov et al. 1997) and
$\chi_\odot\approx-320\pm9^\circ$ (Mishurov and Zenina 1999) were
found from relatively old Cepheids with periods $P<9^d$. Thus, we
get the impression that the masers are intermediate between OB
stars and Cepheids in this parameter. The fact that the kinematics
of these recently formed stars reflects considerably earlier
stages in the motion of regions of active star formation can be a
possible explanation. Testing this assumption requires a separate
study using reliable data on stellar ages.

\section*{CONCLUSIONS}

We proposed a new method of searching for periodicities in the
residual velocities of Galactic objects to estimate the parameters
describing the Galactic spiral density wave in accordance with the
theory of Lin and Shu (1964). In contrast to the method of
harmonic analysis of series that we used previously, this method
based on a periodogram Fourier analysis takes into account the
logarithmic pattern of the Galactic spiral structure and the
position angles of Galactic objects. This allows an accurate
analysis of the velocities for objects distributed in a wide range
of Galactocentric distances to be performed. To increase the
significance of the extraction of periodic signals from data
series with large gaps, we developed a spectrum reconstruction
method based on the generalized maximum entropy method (Bajkova
1992).

Using the proposed methods, we redetermined the Galactic spiral
density wave parameters from the radial velocities of 44 masers
with known trigonometric parallaxes, proper motions, and
line-of-sight velocities. The following spiral density wave
parameters were obtained: the perturbation amplitude $f_R =
7.7^{+1.7}_{-1.5}$ km s$^{-1}$,the perturbation wavelength
$\lambda=2.2^{+0.4}_{-0.1}$~kpc, the pitch angle of the spiral
density wave
 $i=-5^{+0.2^\circ}_{-0.9^\circ}$,
 and the phase of the Sun in the
spiral density wave $\chi_\odot= -147^{+3^\circ}_{-17^\circ}$.

\subsection*{Acknowledgments}

We are grateful to the referees for their useful remarks that
contributed to an improvement of the paper. This work was
supported by the ``Nonstationary Phenomena in Objects of the
Universe'' Program of the Presidium of the Russian Academy of
Sciences and by the ``Multiwavelength Astrophysical Research''
grant no. NSh--16245.2012.2 from the President of the Russian
Federation.

\bigskip{\bf REFERENCES}\bigskip {\small

1. K. Ando, T. Nagayama, T. Omodaka, et al., Publ. Astron. Soc.
Jpn. 63, 45 (2011).

2. Y. Asaki, S. Deguchi, H. Imai, et al., Astrophys. J. 721, 267
(2010).

3. A.T. Bajkova, Astron. Astrophys. Trans. 1, 313 (1992).

4. V.V. Bobylev and A.T. Bajkova, Mon. Not. R. Astron. Soc. 408,
1788 (2010).

5. V.V. Bobylev and A.T. Bajkova, Astron. Lett. 37, 526 (2011).

6. V.V. Bobylev, A.T. Bajkova, and A. S. Stepanishchev, Astron.
Lett. 34, 515 (2008).

7. J. Bovy, D. W. Hogg, and H.-W. Rix, Astrophys. J. 704, 1704
(2009).

8. J. Byl and M.W. Ovenden, Astrophys. J. 225, 496 (1978).

9. D.P. Clemens, Astrophys. J. 295, 422 (1985).

10. S. Dzib, L. Loinard, L.F. Rodriguez, et al., Astrophys. J.
733, 71 (2011).

11. B.R. Frieden and A.T. Bajkova, Appl. Opt. 33, 219 (1994).

12. C. Goddi, E.M.L. Humphreys, L.J. Greenhill, et al., Astrophys.
J. 728, 15 (2011).

13. K. Hachisuka, A. Brunthaler, K.M. Menten, et al., Astrophys.
J. 696, 1981 (2009).

14. T. Hirota, T. Bushimata, Y.K. Choi, et al., Publ. Astron. Soc.
Jpn. 59, 897 (2007).

15. T. Hirota, T. Bushimata, Y.K. Choi, et al., Publ. Astron. Soc.
Jpn. 60, 37 (2008a).

16. T. Hirota, K. Ando, T. Bushimata, et al., Publ. Astron. Soc.
Jpn. 60, 961 (2008b).

17. T. Hirota, M. Honma, H. Imai, et al., Publ. Astron. Soc. Jpn.
63, 1 (2011).

18. M. Honma, T. Bushimata, and Y. Choi, Publ. Astron. Soc. Jpn.
59, 889 (2007).

19. H. Imai, K. Nakashima, T. Bushimata, et al., Publ. Astron.
Soc. Jpn. 59, 1107 (2007).

20. E.T. Jaynes, IEEE Trans. Syst. Sci. Cybern. 4, 227 (1968).

21. M. K. Kim, T. Hirota, M. Honma, et al., Publ. Astron. Soc.
Jpn. 60, 991 (2008).

22. C.C. Lin and F.H. Shu, Astrophys. J. 140, 646 (1964).

23. P.J. McMillan and J.J. Binney, Mon. Not. R. Astron. Soc. 402,
934 (2010).

24. A.M. Mel'nik and A.K. Dambis, Mon. Not. R. Astron. Soc. 400,
518 (2009).

25. A.M. Mel'nik, A.K. Dambis, and A.S. Rastorguev, Astron. Lett.
27, 521 (2001).

26. K.M. Menten, M.J. Reid, J. Forbrich, et al., Astron.
Astrophys. 474, 515 (2007).

27. Yu. N. Mishurov and I.A. Zenina, Astron. Astrophys. 341, 81
(1999).

28. Yu. N. Mishurov, E.D. Pavlovskaya, and A.A. Suchkov, Sov.
Astron. 23, 147 (1979).

29. Yu. N. Mishurov, I. A. Zenina, A.K. Dambis, et al., Astron.
Astrophys. 323, 775 (1997).

30. G.A. Moellenbrock, M.J. Claussen, and W.M. Goss, Astrophys. J.
694, 192 (2009).

31. L. Moscadelli, M. J. Reid, K.M. Menten, et al., Astrophys. J.
693, 406 (2009).

32. K. Motogi, K. Sorai, A. Habe, et al., Publ. Astron. Soc. Jpn.
63, 31 (2011).

33. T. Nagayama, T.Omodaka, A. Nakagawa, et al., Publ. Astron.
Soc. Jpn. 63, 23 (2011).

34. K. Niinuma, T. Nagayama, T. Hirota, et al., Publ. Astron. Soc.
Jpn. 63, 9 (2011).

35. C.S. Oh, H. Kobayashi, M. Honma, et al., Publ. Astron. Soc.
Jpn. 62, 101 (2010).

36. M.J. Reid, K.M. Menten, X.W. Zheng, et al., Astrophys. J. 700,
137 (2009).

37. K. Rohlfs, Lectures on Density Wave Theory (Springer-Verlag,
Berlin, 1977).

38. K.L.J. Rygl, A. Brunthaler, M.J. Reid, et al., Astron.
Astrophys. 511, A2 (2010).

39. K.L.J. Rygl, A. Brunthaler, A. Sanna, et al., arXiv:1111.7023
(2011).

40. K.M. Sandstrom, J.E.G. Peek, G.C. Bower, et al., Astrophys. J.
667, 1161 (2007).

41. A. Sanna, M.J. Reid, T.M. Dame, et al., Astrophys. J. 745, 82
(2012).

42. M. Sato, T. Hirota, M.J. Reid, et al., Publ. Astron. Soc. Jpn.
62, 287 (2010a).

43. M. Sato, M.J. Reid, and A. Brunthaler, Astrophys. J. 720, 1055
(2010b).

44. S. Shiozaki, H. Imai, D. Tafoya, et al., Publ. Astron. Soc.
Jpn. 63, 1219 (2011).

45. A.S. Stepanishchev and V.V. Bobylev, Astron. Lett. 37, 254
(2011).

46. Y.-N. Su, S.-Y. Liu, H.-R. Chen, et al., Astrophys. J. 744,
L26 (2012).

47. V.V. Vityazev, Analysis of Nonuniform Time Series (SPb. Gos.
Univ., St.-Petersburg, 2001) [in Russian].

48. Y.Xu, M.J. Reid, X.W. Zheng, et al., Science 311, 54 (2006).

49. Y.Xu, L. Moscadelli, M.J. Reid, et al., Astrophys. J. 733, 25
(2011).

50. M.V. Zabolotskikh, A.S. Rastorguev, and A.K. Dambis, Astron.
Lett. 28, 454 (2002).

51. B. Zhang, M.J. Reid, K.M. Menten, et al., Astrophys. J. 744,
23 (2012).

}

\end{document}